\documentstyle[11pt,newpasp,twoside,epsf]{article}
\def\Lsun{L$_\odot$}
\def\Msun{M$_\odot$}
\def\Mh{M_{\rm h}}

\def\Macc{\dot{M}_h^{\rm acc}}
\def\Ha{H$_\alpha$}
\def\gtrsim{\,\,\raise0.14em\hbox{$>$}\kern-0.76em\lower0.28em\hbox  
{$\sim$}\,\,}  
\def\lesssim{\,\,\raise0.14em\hbox{$<$}\kern-0.76em\lower0.28em\hbox  
{$\sim$}\,\,}

\def\edcomment#1{\iffalse\marginpar{\raggedright\sl#1\/}\else\relax\fi}
\reversemarginpar

\begin{document}
\title{The link between symbiotic stars and chemically-peculiar red giants}
 \author{Alain Jorissen}
\affil{Institut d'Astronomie et d'Astrophysique, Universit\'e Libre de
  Bruxelles, C.P. 226, Boulevard du Triomphe, B-1050 Bruxelles, Belgium}

\begin{abstract}
Barium stars and technetium-poor, extrinsic S stars are binary systems 
with a white dwarf companion, and with orbital elements
similar to those of symbiotic systems. One may thus wonder whether
these various families of binary systems involving red giant stars are 
somehow related. This question is actually twofold: \\
(i) Do barium and binary S stars exhibit some symbiotic activity?\\
(ii) Do symbiotic systems exhibit overabundances of s-process elements
like barium and S stars? \\
This paper reviews the current situation regarding these two questions.
\end{abstract}

\section{The zoo of red giant stars}
\label{Sect:zoo}

Symbiotic stars (SyS), barium stars and technetium-poor S stars are
three families involving
giant stars where binarity plays a key role.
Thanks to the progress of UV astronomy, the binary nature of SyS
now goes undisputed as their UV spectra bear the signature of a hot
component, generally a white dwarf
(WD), 
whereas the optical spectrum is dominated by the red giant 
(Miko\l ajewska, this conference).

Barium stars, first identified by Bidelman \& Keenan (1951),
are G and K giants where carbon and elements heavier than Fe, 
like Ba and Sr, have surface abundances in excess of the solar value 
(e.g. Wallerstein at al. 1997 and references therein),
i.e., [X/Fe] $ = \log \left[\left(N({\rm X})/N({\rm Fe})\right) / \left(N_\odot({\rm X})/N_\odot({\rm Fe})\right)\right] > 0$, where $N(\rm X)$ stands for the abundance of
element X. 
Heavy elements like Sr and Ba are synthesized by the so-called {\it
  s-process} of nucleosynthesis, a sequence of neutron captures 
starting on abundant  seed nuclei like iron-group
elements (Burbidge et al. 1957;
Wallerstein et al. 1997). The operation of the s-process is
commonly associated with thermal pulses (TPs) occurring on the asymptotic
giant branch (AGB; e.g. Goriely \& Mowlavi 2000). 
AGB stars have a complex internal structure, consisting of a
carbon-oxygen core, helium- and hydrogen-burning shells and a
deep convective envelope (Olofsson \& Habing 2003). TPs 
are a recurrent thermal instability affecting the He-burning shell. 
Right after a TP, the inner boundary of the convective envelope
may penetrate the intershell region where the s-process operated.
As a result of this mixing process (the so-called `third dredge-up' -- 3DUP), 
s-process-enriched material is brought to the stellar surface. 
Barium stars are, however, too warm and of too low
a luminosity to be thermally-pulsing AGB (TP-AGB) stars (Scalo 1976).  
With the discovery that the barium stars are all single-line spectroscopic
binaries (McClure et al. 1980; McClure 1983; Jorissen et al. 1998),  
their chemical
peculiarities have been ascribed to mass transfer across the binary system. 
The unseen companion is almost certainly a WD
(McClure \& Woodsworth 1990). In a former state of the binary system,
that WD was a TP-AGB 
star where the s-process and the 3DUP were operating. If 
mass transfer processes (like wind accretion or Roche lobe overflow -- 
RLOF)
were able at that stage to dump s-process-rich and C-rich
material from the AGB star onto its companion, that companion would be
turned into a barium star (see Jorissen 2003 for a review). This is
the so-called {\it binary paradigm} to account for the {\it barium
  syndrome}.

The stars of spectral type S (first identified by Merrill in 1922) 
exhibit chemical peculiarities very
similar to those of barium stars. 
However, the S star family hosts two
kinds of stars, the so-called {\it extrinsic} and {\it intrinsic} S stars,
having very different evolutionary status (Van Eck \& Jorissen 2000). 
They are best distinguished
by the presence or absence of Tc lines, a heavy element with no stable
isotopes (Van Eck \& Jorissen 1999). 
Intrinsic S stars exhibit Tc lines, and  are AGB
stars where the s-process is operating, as indicated above. There is
thus no need to invoke mass transfer across a binary system to
account for their chemical peculiarities.
On the contrary, extrinsic S stars have no Tc lines, and are all 
binary stars (Jorissen et al. 1998). 
They are the cool analogs of
barium stars. For both barium and extrinsic S stars 
(collectively referred to as
peculiar red giants -- PRG -- in the following), the companion mass 
inferred from the 
observed mass-function distribution is consistent with that companion
being a WD (McClure \& Woodsworth 1990; 
Jorissen et al. 1998; North et al. 2000). 

Since both PRG and SyS are binary systems
consisting of a red giant and a WD, the relation between these
families ought to be elucidated; more precisely:
(i) Do PRG exhibit symbiotic activity? 
(ii) Do SyS exhibit the barium syndrome?

\section{Do PRG exhibit symbiotic activity?}

\subsection{Physical conditions required to trigger symbiotic activity}
\label{Sect:formulae}

Before reviewing the symbiotic activity observed in the various
classes of PRG (Sect.~\ref{Sect:PRGsymbio}), it is useful to formally identify
the physical conditions required to trigger symbiotic activity. 
The key to this activity lies in the luminosity of a hot ($T >
50\;000$~K) companion
star, estimated to be at least 10~\Lsun\ (M\"urset et al. 1991; 
Yungelson et al. 1995), which emits UV radiation that ionizes the
wind from the cool star, giving rise to the rich emission-line
spectrum  (Nussbaumer \& Vogel 1987). 
Different physical processes, all related to the accretion of mass by
the companion, may be at the origin of this high
luminosity (Yungelson et al. 1995; Iben \& Tutukov 1996): 
(i) steady hydrogen burning at the surface of a WD
(ii) thermonuclear flashes at the surface of a WD [associated with
symbiotic novae] 
(iii) release of gravitational energy associated with accretion, 
partially converted into
radiative energy. Regimes (i) and (ii) correspond to accretion rates
respectively above and below
some critical value $\dot{M}^{\rm acc}_{\rm crit}$ 
(Eq.~b of Table~\ref{Tab:formulae},
about $5\times 10^{-8}$~M$_\odot$~y$^{-1}$ for a 0.6~\Msun\ WD). 
Regime (iii) applies to main-sequence or neutron-star accretors, and
to WD accretors in between H-burning flashes (since the accretion
luminosity is smaller than the H-burning luminosity
at any given WD mass $M_h$; see Table~\ref{Tab:formulae} and 
Fig.~5 of Iben \& Tutukov 1996). Formulae relating the luminosity
$L_h$ of
the compact companion to its mass $M_h$ for these three regimes are provided 
in Table~\ref{Tab:formulae}.
A convenient analytical formula for the accretion rate $\Macc$ is
available in the framework of the Bondi-Hoyle regime  of wind
accretion  (Bondi \& Hoyle 1944), although
detailed hydrodynamical simulations 
(Theuns, Boffin \& Jorissen 1996; Mastrodemos \& Morris 1998; Folini
\& Walder 2000)
have shown that the Bondi-Hoyle accretion rates (Eq.~c  in
Table~\ref{Tab:formulae} with $\eta \sim 1$) are generally 
an order of magnitude too large when the wind velocity is of the
same order as the orbital velocity, as it is the case for SyS.  
The parameter $\eta$ appearing in Eq.~(c) of
Table~\ref{Tab:formulae} must therefore be taken of the order of 0.1
to reflect the results of numerical simulations when $v_{\rm wind} /
v_{\rm orb} \lesssim 1$.

To summarize, Table~\ref{Tab:formulae} provides relationships that
may in principle be
used to propagate the condition $L_h > 10$~\Lsun\ defining a SyS into
constraints on the various physical parameters involved (the hot
companion mass $M_h$,
the cool star mass $M_c$, its luminosity
$L_c$ and temperature $T_c$, 
the metallicity $Z$ and the orbital period $P$),
namely:
\begin{eqnarray*}
10\; {\rm L}_\odot < L_h & = & f\;[M_h, \Macc]\\ 
                         & = & f\;[M_h, \Macc(\dot{M}_c^{\rm
                               wind},v_{\rm wind},M_c,P)]\\ 
  & = & f\left[M_h, \Macc\left(\dot{M}_c^{\rm wind} [ M_c, T_c, L_c
( M_c, T_c, Z )], v_{\rm wind}(L_c,Z),M_c,P)\right)\right]\\
  & = & f\;[M_h, M_c, T_c, Z, P].  
\end{eqnarray*}

The metallicity $Z$ enters the discussion through the mass-loss properties 
of the cool star, especially its wind velocity (Eq.~e of
Table~\ref{Tab:formulae}; Van Loon 2000). 
In the various empirical parametrizations reviewed
by Zijlstra (1995), the wind mass loss rate $\dot{M}_c^{\rm wind}$ 
depends explicitely upon $M_c$, $L_c$ and the stellar radius $R_c$, 
which transforms into a function of $M_c, T_c$ and 
$Z$ using the Stefan-Boltzmann formula to eliminate $R_c$ and the
evolutionary track to express $L_c$ as a function of $M_c, T_c$ and 
$Z$. All the empirical formulae reviewed by Zijlstra (1995) predict
that {\it the mass loss rate increases with increasing $L_c$ and $R_c$}. 
This will turn out to be a very important property to understand the
occurrence of -- or lack of -- symbiotic activity in the various
families of PRG.

The above discussion assumes that the system is detached, which sets
yet another constraint, namely $P > P_{\rm RLOF}(R_c,M_c,M_h)$ (Eq.~d of
Table~\ref{Tab:formulae}) where $P_{\rm RLOF}(R_c,M_c,M_h)$ is the
orbital period of the (semi-detached) system with a cool star of
radius $R_c$ filling its Roche lobe.

\begin{table}
\caption[]{\label{Tab:formulae}
A compendium of formulae to transform the condition $L_h >
10$~\Lsun\ defining SyS into constraints on the various 
physical parameters involved. Luminosities, masses and radii are
expressed in solar units, mass loss and accretion rates in \Msun~y$^{-1}$, 
velocities in km~s$^{-1}$ and periods in y. Subscripts $h$ and $c$
refer to the hot and cool components, respectively
}

\vspace{5mm}
\fbox{\parbox{14cm}{
$
\begin{array}{lllll}
\noalign{\bf $\bullet$ Regime (i): steady H-burning on WD:
\boldmath $\dot{M}_h^{\rm acc} >   \dot{M}_{\rm crit}^{\rm acc}$}\\

 L_h^{\rm He} & = 4\; 10^5 M_h^{6.5}   & \hspace{\fill} (M_h <
   0.5:  {\rm RGB-like,\; He\; core)}  & \tt{(a)}
\smallskip\\
 L_h^{\rm CO} & = 60\;000\; (\Mh - 0.52) & \hspace{\fill} (M_h >
   0.5: {\rm AGB-like,\; CO\; core)}  & \tt{(a)}\cr   
\medskip\\

\noalign{\bf $\bullet$ Regime (ii): H-flash on WD: 
\boldmath $\dot{M}_h^{\rm acc} <  \dot{M}_{\rm crit}^{\rm acc}$ } \cr

 L_h^{\rm cold\; He} & = 46\;000\; (\Mh - 0.26) & \hspace{\fill} \equiv L {\rm
   (plateau \; after\; flash\; on\; a\; cold\; He\; core)} & \tt{(a)} \cr   
\medskip\\

\noalign{\bf $\bullet$ Regime (iii): accretion luminosity\hspace{\fill}(main
sequence companion,}
\noalign{\bf \hspace*{\fill} or between H-flashes on a WD)}\cr
L_h^{\rm acc} & = 3\;10^7 \Macc\;\frac{M_h}{R_h} & \hspace{\fill} {\rm (general\; expression)}\cr 
             & = 3\;10^9 \Macc \frac{M_h}{1.96 - 1.16 M_h} &
             \hspace{\fill} {\rm  (WD)}  & \tt{(a)} \cr
\end{array}
$
\medskip\\
\hspace*{5mm} {\rm with} $L_h^{\rm acc}(\dot{M}^{\rm acc}_{\rm
    crit}) << L_h^{\rm He}, L_h^{\rm CO} < L_h^{\rm cold\; He}$\\
}}
\medskip\\
{\bf The operation of a given regime is dictated by the ratio 
\boldmath $\dot{M}_h^{\rm acc}/\dot{M}_{\rm crit}^{\rm acc}$:}\\
where
\smallskip\\
\hspace*{5mm} $\dot{M}_{\rm crit}^{\rm acc}= 10^{-9.31 + 4.12 M_h - 1.42
M_h^2}$ \hspace{\fill} \tt{(b)}\\

\hspace*{5mm} $\dot{M}_h^{\rm acc} = -\dot{M}_c^{\rm wind}
\;
\eta \; \mu^2 \; \frac{k^4}  {\left[ 1 + k^2 + \left(\frac{c}{v_{\rm
wind}}\right)^2\right]^{3/2}}$ \hspace{\fill}  \tt{(c)}

\hspace*{10mm} {\rm where}\\
\hspace*{15mm} $\mu = M_h/(M_h + M_c)$\\
\hspace*{15mm} $k = v_{\rm orb} / v_{\rm wind} = 30 \; \left(\frac{M_h + M_c}{P}\right)^{1/3} \; / \; v_{\rm wind}$ 
\medskip\\
\hspace*{15mm} $P > P_{\rm RLOF} [R_c, M_c, M_h ] =
\frac{3\times 10^{-4}\;R_c^{3/2}}{(M_c + M_h)^{1/2} (0.38 + 0.2 \log
  (M_c/M_h))^{3/2}}$
\hspace{\fill}\tt{(d)}\bigskip\\ 

$
\begin{array}{llll}
\hspace*{15mm} \eta \sim & 1   & {\rm if}\; k^{-1} = v_{\rm wind} / v_{\rm orb} >> 1 & \hspace{\fill} {\rm  (Bondi-Hoyle\; regime)}\\
\hspace*{15mm} \eta \sim & 0.1 & {\rm if}\; k^{-1} = v_{\rm wind} / v_{\rm orb} \le 1
&\\
\end{array}
$
\bigskip\\

\hspace*{15mm} $c =$ {\rm sound velocity}\\
\hspace*{15mm} $v_{\rm wind} \propto L_c^{1/4} Z^{1/2}$
\hspace{\fill}\tt{(e)}\smallskip\\ 
\hspace*{15mm} $\dot{M}_c^{\rm wind} = \dot{M}_c^{\rm wind} [ M_c, T_c, L_c
( M_c, T_c, Z ) ]$
\hspace{\fill}\tt{(f)}\bigskip\\

References: \tt{(a) Iben \& Tutukov (1996) (b) Yungelson et al. (1995)
(c) Theuns et al. (1996) (d) Jorissen (2003) 
(e) Van Loon (2000) (f) Zijlstra (1995)}

\end{table}

\subsection{Symbiotic activity among PRG}
\label{Sect:PRGsymbio}

\subsubsection{Ba stars.}

Symbiotic activity among barium stars is basically inexistent, except
for the barium supergiant 56~Peg and for HD 46407, a barium star with
one of the shortest orbital periods (456.6~d). 
56~Peg is an X-ray source with a
hot WD (Schindler et al. 1982; Dominy \& Lambert 1983; Schwope et
al. 2000).    
HD 46407 exhibits dust obscuration episodes (Jorissen
1994, 1997) reminiscent of 
those observed in symbiotic Miras (Munari \& Whitelock 1989),
although to a much lesser extent.
This quasi-absence of symbiotic activity among barium stars 
is not surprising given their rather
low mass-loss rates $\dot{M}_c^{\rm wind}$ (Drake,
Simon, \& Linsky 1987), consistent with their luminosities of RGB
(rather than AGB) stars (Scalo 1976).
Equation~(c) of
Table~\ref{Tab:formulae} then indicates that the accretion rate 
by the companion will be low as well 
($< 10^{-10}$~M$_\odot$~y$^{-1}$; Jorissen 1997). If anything, 
{\it this situation
leads to H-flashes on the WD companion (regime ii)}, 
although no such events have yet been reported for barium stars.

\subsubsection{S stars.}

Van Eck \& Jorissen (2002; their Table~2) 
have collected all S stars where signatures of
symbiotic activity have been reported. All of these -- with the
exception of the Henize S stars (see below) -- result from
serendipitous discoveries, and thus rely on different diagnostics
of symbiotic activity which are not equally sensitive to
$\dot{M}_h^{\rm acc}$. For instance, \Ha\  emission is not 
observed in the long-period system HD 49368 (=V613~Mon; $P \sim 3000$~d)
despite a strong UV excess (Ake 1996, priv. comm.). 
A similar situation is encountered for the SyS EG~And (see in
particular the discussion in Sect.~4.4 of Munari 1994).

\begin{figure}
\plotfiddle{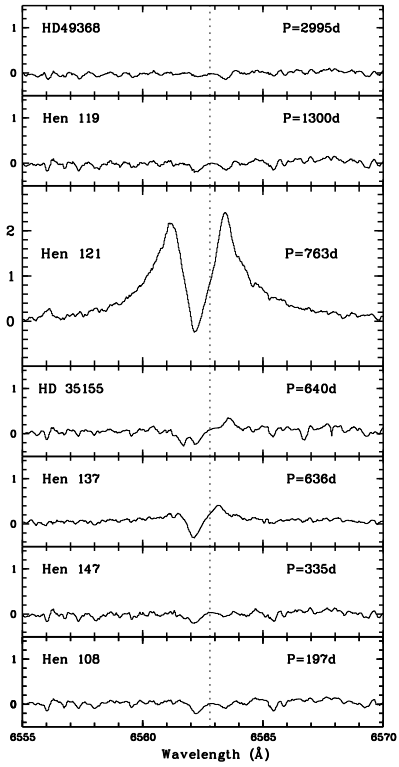}{10cm}{0}{60}{60}{-250}{0}
\caption[]{\label{Fig:HaP}
Residual \Ha\ profiles obtained  after subtracting a pure absorption
\protect\Ha\ profile (from the S star Hen4-6), for
S stars with known orbital periods (as indicated in the upper right
corner), decreasing from top to bottom.
It is clearly apparent that S stars with \Ha\ emission are only found in
the period range 600 -- 800~d. The dashed vertical line is the rest
\Ha\ wavelength
}
\end{figure}

Therefore, to find any systematics (like correlation
with orbital period) 
requires a more
systematic approach. The 66 binary S stars from the Henize sample (Van 
Eck \& Jorissen 2000) offers such an opportunity. These stars were searched for
\Ha\ emission, resulting in the discovery 
of two new SyS (Hen4-18 and Hen4-121, following the
SIMBAD terminology) and of two marginal cases (Hen4-134 and Hen4-137; Van
Eck \& Jorissen 2002). Their \Ha\ profiles, displayed in Fig.~\ref{Fig:HaP},
are typical of SyS, 
since they closely
resemble those labelled `S-3' by Van Winckel, Duerbeck \&
Schwarz (1993; see Lee 2000 for a discussion of the formation
mechanism of the
\Ha\ emission line in SyS).  Figure~\ref{Fig:HaP} reveals that
symbiotic S stars
with \Ha\ emission are found in the narrow period range 600 -- 800~d
(as noted above, S stars with longer periods may 
exhibit other signatures of symbiotic activity, though).

\begin{figure}
\plotone{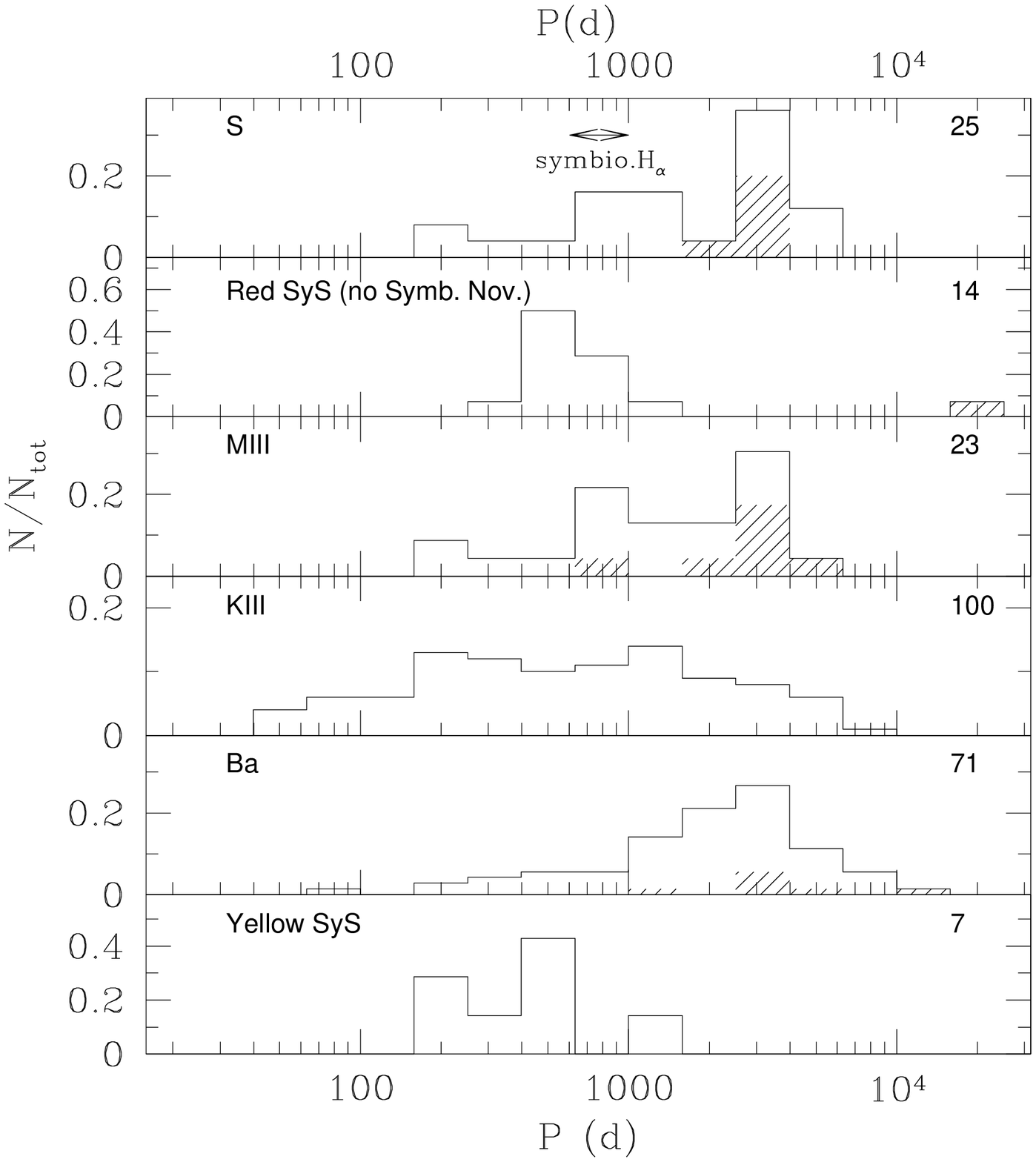}
\caption[]{\label{Fig:P}
Comparison of the period distributions for samples of binary systems
with different kinds of red
giant primaries: S stars (Jorissen et al. 1998), red
SyS  (excluding symbiotic novae and symbiotic Miras; M\"urset \& 
Schmid 1999), M giants 
(Jorissen et al., in preparation) and 
K giants (Mermilliod 1996). The lower two
panels  present the orbital-period distribution for barium stars
(Jorissen et al. 1998) and
yellow SyS (M\"urset \& Schmid 1999).
The shaded area marks stars with only a lower limit available on their orbital
period. The numbers in the upper right corner 
of each panel correspond to the sample size  
}
\end{figure}

The absence of \Ha\ emission 
among the shortest-period systems 
is an interesting result, which confirms that the
orbital period is not the primary parameter controlling symbiotic
activity, as inferred from Table~\ref{Tab:formulae}. 
The key parameter is rather the accretion rate $\dot{M}_h^{\rm acc}$, 
which is a combination of several parameters, including  
$P$ and $\dot{M}_c^{\rm wind}$ (Eq.~c of
Table~\ref{Tab:formulae}). The absence of symbiotic activity among
short-period S stars is likely due to their low mass-loss rates 
$\dot{M}_c^{\rm wind}$, which are in turn a consequence of the fact
that S stars with short orbital periods 
cannot be located very far up the giant branch.
They must thus have smaller radii, luminosities, and hence mass loss rates,
than S stars with longer orbital periods. 
The orbital period indeed imposes a maximum admissible radius $R_c$
corresponding to the critical Roche radius (Eq.~d of
Table~\ref{Tab:formulae}). Put differently, this condition  means
that to any given spectral type corresponds a minimum admissible orbital 
period for unevolved (i.e., pre-mass-transfer) {\it detached} systems.
Such a correlation is 
clearly apparent on Fig.~\ref{Fig:P}, which shows that 
the minimum orbital period among K
giants is 40~d,  increasing to 200~d for M giants, and even to 400~d
for giants later than M0~III. 
A similar correlation has 
been found by M\"urset \& Schmid (1999) among SyS (see
also Harries \& Howarth 2000), despite the fact that SyS are not
really unevolved systems.

Finally, it must be noted that the period range of red SyS
(excluding symbiotic novae and symbiotic Miras) matches fairly well 
that of symbiotic S stars with \Ha\ emission and, moreover, corresponds to the
short-period tail of the M~III binaries. The same holds true for 
the period distribution of yellow SyS, which corresponds to the 
short-period tail of barium stars ({\it but not} to the short-period
tail of K~III binaries!).  
The reason for this is clear: barium stars and yellow 
SyS are two families which involve WD companions, unlike K~III
binaries which may also involve main sequence companions. Therefore,
barium and yellow SyS share the same evolutionary history, 
namely one component has gone through the AGB. 
The fact that 
these systems once contained a very extended star sets a lower
limit on their orbital period, as discussed above in relation with Eq.~d 
(for a more detailed 
discussion of these aspects, see Jorissen 2003).
But this restriction does not apply to the sample of K~III binaries, 
hence they may contain systems with much shorter orbital periods than
barium stars and yellow SyS.

\begin{table}
\caption[]{\label{Tab:yellow}
The barium syndrome among yellow SyS. The spectral type of the
cool component is taken from M\"urset \& Schmid (1999), or references therein.
In column labelled `nebula', `y' means that an optical nebula has been 
detected, and `PN' that, based on its emission line
spectrum, the star has traditionally been
included in planetary nebulae catalogues,
even though no optical nebula may be visible}

\tabcolsep 2pt
\begin{tabular}{lllrrcllll}
\medskip\\
\hline
\smallskip\\
 Name          & Sp. Typ. & [Fe/H]   & \multicolumn{1}{c}{$V_r$}    &
\multicolumn{1}{c}{$b$}   & [Ba/Fe] &
$V \sin \; i$ & nebula & Ref.\\
               &          &          & \multicolumn{1}{c}{(km/s)}   
                                         & \multicolumn{1}{c}{($^\circ$)}
&& \multicolumn{1}{c}{(km/s)}\\
\tableline
\medskip\\
\noalign{\hspace*{\fill}d'-type\hspace*{\fill}}\\
\hline\\
V417 Cen & G8-K2 & $\sim 0.0$ & & $-1$ & 0.5 &  70 & y &  (5,11)\\
HDE 330036& G5 & $\sim 0.0$ & $-14$ & $+4$ & 0.6 & 100 & PN & (5,14)\\
\multicolumn{1}{r}{=Cn 1-1}\cr 
AS 201  & G5 & $\sim 0.0$ & & $+7$ & 0.4 &  25 & y &  (5,12)\\
V471/V741 Per & G5 &  ? &$-$12 & $-$9 & $> 0$ & &PN & (2)\\ 
\multicolumn{1}{r}{=M 1-2}\cr 
St H$_\alpha$ 190 & G5 & 0.0 & $\sim 10$ & $-$35 & $\sim 0.5$ & 100 &
bip. outf.& (10,13)\\ 
Wray 157 & G5 & ? \\ 
Hen 1591 & $<$ K4 & ?\\
\hline\\
\noalign{\hspace*{\fill}s-type\hspace*{\fill}}\\
\hline\\
UKS Ce-1 & C4,5Jch & ? & +20 & +20 & $>0$ &&&  (6)\\
S 32  & C1,1CH & ? & +325 & $-$30 &  $>0$ &  &&(6,14)\\
Hen 2-467 & K0  & -1.1 & $-$109 & $-$12 &  +0.8  &&n & (4,16)\\ 
BD-21:3873 & K2 & -1.1 & +204 & +37 & +0.5 & & n &(3,15,16) \\
           &     & -1.3   &          &        & +0.3 & &&(9)\\  
AG Dra & K2 & -1.3 & $-$148 & +41 & +0.5 & & n &(8,16)\\
CD -43:14304 & K7 &?   & +27 & $-41$ & ?  &  &&(7)\\
\hline
\medskip\\
\noalign{\hspace*{\fill}Chemical evolution of the Galaxy\hspace*{\fill}}\\
   & & -1.0  & & &   $< 0.2$ &&&  (1) \medskip\\
\tableline
\tableline
\end{tabular}
\vspace{3mm}\\
{\small References: 
(1) Edvardsson et al., 1993, A\&A, 275, 101 (2) Grauer \& Bond, 1981, PASP,
93, 630 (3) Pereira et al., 1997, AJ, 114, 2128 (4) Pereira et al., 1998, AJ, 116,
1977 (5) Pereira et al., 2003, this conference
(6) Schmid, 1994, A\&A, 284, 156 (7) Schmid et al., 1998, A\&A, 329, 986 
(8) Smith et al., 1996, A\&A, 315,
179 (9) Smith et al., 1997, A\&A, 324, 97 (10) Smith et al., 2001, ApJ, 556,
L55 (11) Van Winckel et al., 1994, A\&A, 285, 241  (12) Schwarz, 1991, A\&A,
243, 469 (13) Munari et al., 2001, A\&A, 369, L1 
(14) Schmid \& Nussbaumer, 1993, A\&A, 268, 159
(15) Munari \& Patat, 1993, A\&A, 277, 195 
(16) Corradi et al., 1999, A\&A, 343, 841}
\end{table}

\section{Do SyS exhibit the barium syndrome?}

\subsection{Physical conditions required to trigger the s-process}
\label{Sect:s-process}

Here again, we start with a formal discussion of the conditions 
required to trigger the operation of the s-process in AGB stars,
before considering the question whether SyS exhibit the
barium syndrome.
These conditions are:\\ 
\begin{displaymath}
\fbox{$
\begin{array}{lll}
M_h & > & 0.5\; {\rm M}_\odot\\
Z & < & Z_\odot
\end{array}
$}
\end{displaymath}

The first condition on the core mass of the AGB star (which is
identical to the mass $M_h$ of the WD in the present SyS) expresses the fact
that the  AGB star must have gone through the TP phase
(Sect.~\ref{Sect:zoo}). Wagenhuber \& Groenewegen (1998, their Fig.~7)
provide the AGB core mass at the first TP,  for AGB stars
of various metallicities and initial masses, from which the first
condition is derived.

The second condition, expressing that the efficiency of the s-process
is higher in low-metallicity AGB stars, was first suggested by Clayton (1988).
This efficiency  may be expressed by a single quantity, 
$n_c$, the number of neutrons captured per (iron) seed nuclei.  For example,
$n_c \simeq 138 - 56 = 82$  is required for $^{138}$Ba to be synthesized from
$^{56}$Fe. 
Assuming that there are no strong neutron poisons, all neutrons will
be captured by Fe and its daughter nuclei, so that
$n_c = N({\rm neutron\; supply})/N({\rm
  Fe})$. Neutrons are supplied by a `neutron source', namely 
$^{13}$C($\alpha$,n)$^{16}$O as it is currently believed for AGB stars 
(see e.g., Wallerstein et al. 1997), with $^{13}$C resulting from the
so-called proton-mixing scenario (e.g., Goriely \& Mowlavi 2000). 
In this scenario, protons from the convective envelope are mixed
in layers enriched in $^{12}$C by the former TP, resulting
in the synthesis of $^{13}$C through the chain 
$^{12}$C(p,$\gamma)^{13}$N($\beta)^{13}$C. In the framework of the
proton-mixing scenario, $^{13}$C may be considered as `primary'
(in the sense of galactic chemical evolution), 
since it is synthesized from primary species, namely hydrogen from the 
envelope and $^{12}$C resulting from the $3\alpha$ reaction. Assuming
that there is no leak in the neutron production by the $^{13}$C
source, all available $^{13}$C nuclei will yield neutrons, so that
$n_c = N({\rm neutron\; supply})/N({\rm Fe}) = N(^{13}{\rm C})/N({\rm
  Fe}) \propto 1/Z$. 

This expectation seems to be borne out by empirical evidence; 
see the discussion in Jorissen \& Boffin
(1992) and Jorissen et al. (1998; their Sect.~8)
showing that binarity is not a sufficient condition to form a barium
star, but that a subsolar metallicity seems to be required as well. 

At this point, it should be remarked, however, that 
metallicity is also likely to have an impact on the wind
accretion rate: at high $Z$, the wind velocity is larger (Eq.~e
of Table~\ref{Tab:formulae}; also Zuckerman \& Dyck 1989) and
therefore the accretion rate is smaller, since $\dot{M}_h^{\rm acc}
\propto v_{\rm wind}^{-4}$ approximately.
Therefore, barium stars may not form at high metallicities, not only
because the s-process in the former companion AGB star is less
efficient, but also because accretion of its wind is less efficient. 

Because of this sensitivity upon metallicity, s-type yellow SyS,
d'-type yellow SyS and red
SyS have to be considered separately in the following, since 
they belong to different galactic populations.

\subsection{s-Type yellow SyS are PRG}

\begin{figure}
\plottwo{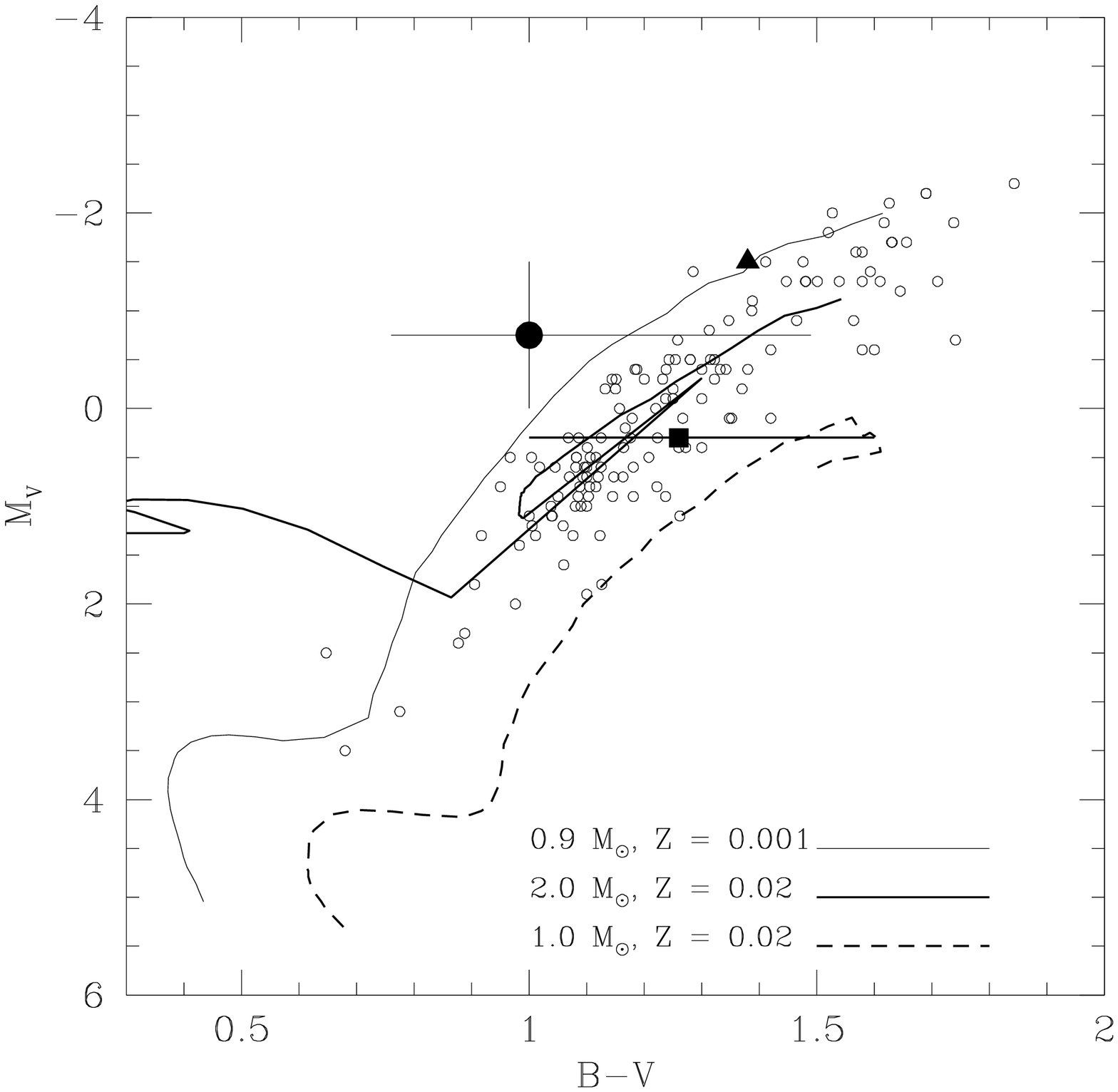}{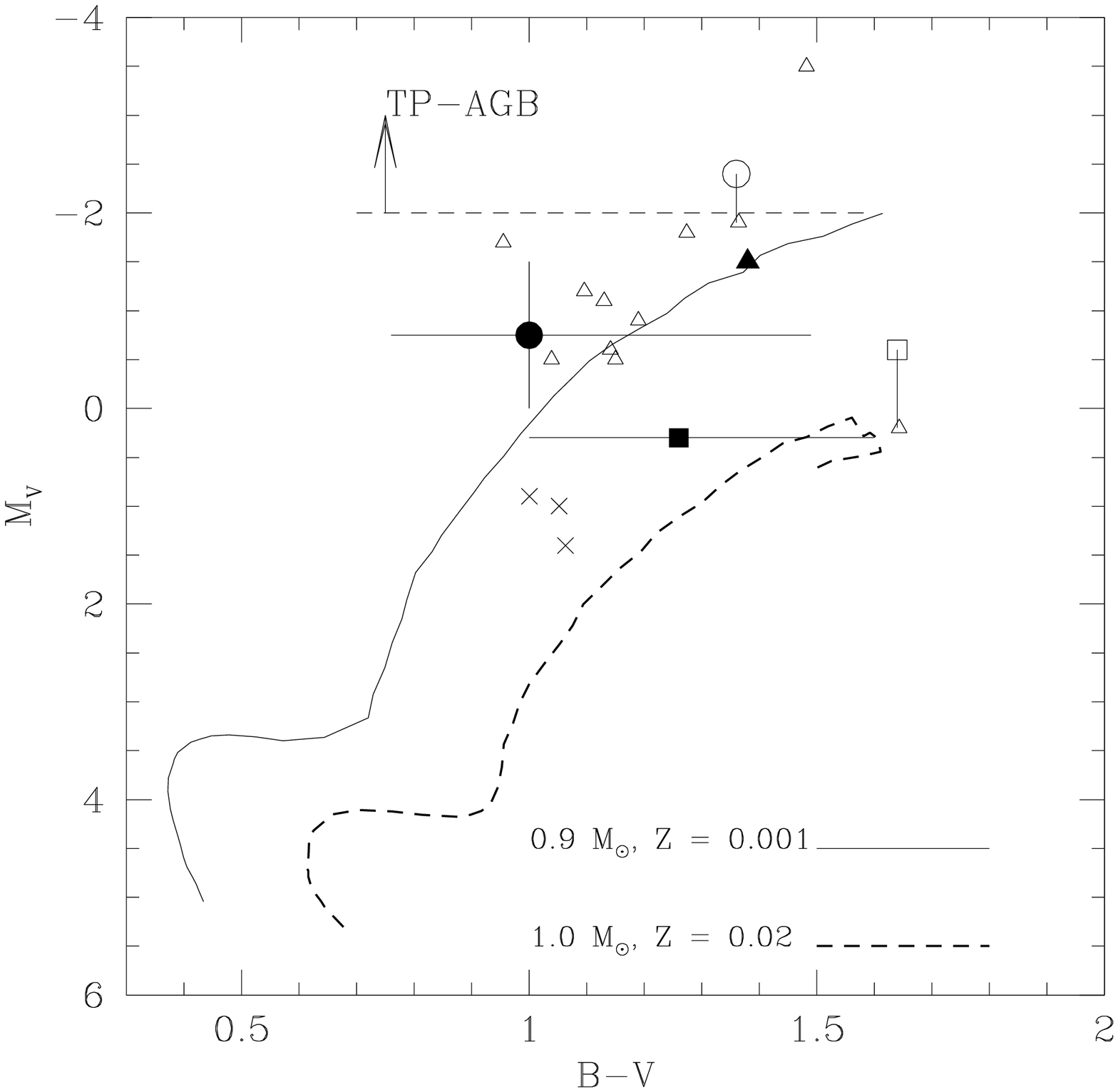}
\caption[]{\label{Fig:mdBa}
Left panel (a): Evolutionary tracks of Schaller et al. (1992) compared
with the locations of classical barium stars (stars labelled `G' in 
Mennessier et al. 1997; open dots) and the yellow SyS AG~Dra
(filled circle), BD $-21:3873$ (filled triangle) and Hen 2-467 (filled 
square). The bolometric magnitudes were taken from the references
listed in Table~\ref{Tab:yellow}. These bolometric magnitudes were combined
with bolometric corrections from Bessel et al. (1998) and 
$B-V$ indices from Munari et al. (1992) and Munari \& Buson (1992)
 to yield the absolute visual magnitudes.\\ 
Right panel (b): Same as the (a) but
for yellow SyS (filled symbols as in the left panel) and metal-deficient
barium stars (open triangles: 
stars flagged as `H' by Mennessier et al. 1997; crosses: CH stars
also flagged as `H' by Mennessier et al. 1997;  
open circle: HD 104340, open square: HD 206983 from Junqueira \&
Pereira 2001). The dashed horizontal line represents the luminosity
at the first TP in a 1~\Msun\ AGB star of metallicity [Fe/H] =$-1.8$
according to Lattanzio (1991)  
}
\end{figure}

All known yellow SyS are listed in Table~\ref{Tab:yellow}, which 
reveals that all the stars
studied so far exhibit the barium syndrome.
Yellow SyS with a {\it stellar} infrared continuum (s-type, as opposed 
to the dusty d'-type; see below) 
are clearly halo objects, as revealed by their low
metallicities and high space velocities (CD $-43:14304$ may be an exception; 
however, it is of spectral type K7, and should perhaps not be included
in the family of yellow SyS). The presence of the barium
syndrome among a family of binary stars belonging to the halo fully supports
the discussion of Sect.~\ref{Sect:s-process} about the conditions required for
s-processing. It should be added at this point that s-type yellow SyS, 
with their metallicities lower than classical barium stars, may be
expected to be, on average, more luminous than the latter (see Fig.~11
of Smith et al. 1996 comparing the luminosity function of Pop.I and
Pop.II K giants).   This is a direct consequence of
the fact that evolutionary tracks
shift towards the blue in the Hertzsprung-Russell (HR) diagram as metallicity
decreases, as shown in Fig.~\ref{Fig:mdBa}b.
Fig.~\ref{Fig:mdBa}a confirms that the yellow
SyS AG~Dra and BD $-21:3873$ are indeed 
more luminous than classical barium stars.
This difference in the average luminosity -- and hence mass-loss rate 
-- of the two populations thus
explains why yellow SyS, despite hosting a K giant, exhibit symbiotic
activity whereas barium stars do not. The larger mass-loss rates for
the cool components of s-type yellow SyS  -- as compared to Ba stars
-- may be inferred from the comparison of their IRAS [12] $-$ [25]
color indices, which reflect the amount of dust present in the
system: ([12] $-$ [25])$_{\rm Ba} < $ 0.1, as compared to 0.45 for
AG~Dra (Smith et al. 1996). M\"urset et al. (1991) and Drake et
al. (1987) provide direct measurements (or upper limits) for the mass
loss rates of AG~Dra and of Ba stars, respectively, which confirm the 
above conclusion.

{\it Metal-deficient barium stars} (with
metallicities in the range $-1.1$ to $-1.8$ comparable to that of
yellow SyS) were identified by Luck \& Bond (1991) and
Mennessier et al. (1997), and occupy the same region of the HR diagram 
as yellow SyS (Fig.~\ref{Fig:mdBa}b). 
The question thus arises why metal-deficient
barium stars are not SyS. Different answers must be seeked, depending 
upon their absolute visual magnitudes $M_{\rm V}$.
The most luminous systems, with $M_{\rm V} < -2$, are likely located
on the TP-AGB\footnote{According to Lattanzio (1991), 
the first TP in a 1~\Msun\ AGB star of  metallicity [Fe/H]$=-1.8$
occurs at $M_{\rm bol} = -3$, corresponding to  $M_{\rm V} \sim -2$},
so that their Ba syndrome may be explained by internal
nucleosynthesis. They ought thus not be binaries, and therefore cannot 
be SyS! HD~104340 (open circle in Fig.~\ref{Fig:mdBa}b), 
a metal-deficient Ba star studied by Junqueira \&
Pereira (2001), provides a good illustration of this situation, since   
it lies above the TP-AGB threshold and 
15 unpublished CORAVEL radial-velocity measurements spanning 7 y do
not reveal any clear orbital motion.

The less luminous and warmest among metal-deficient Ba stars, clumping
around $M_{\rm V} \sim +1$ in the HR diagram, 
are also sometimes 
classified as CH stars (crosses  in Fig.~\ref{Fig:mdBa}b).  
They are not losing mass at a large enough
rate to trigger any symbiotic activity, as revealed by their small
[12] $-$ [25] color indices ($< 0.3$; Smith et al. 1996).

Finally, at intermediate luminosities ($-2 \lesssim M_{\rm V} \lesssim
+1$), metal-deficient Ba stars are not luminous enough to be TP-AGB 
(hence they should be binaries), but yet their mass loss rates must be
large enough to trigger symbiotic activity, since the yellow SyS
belong to the same luminosity range. 
It would thus be of great interest to check (i) the
binary nature of those metal-deficient Ba 
stars\footnote{Primary targets with independent confirmation
of their halo nature are 
HD~5424 (binary with $P = 1881$~d; Jorissen et al. 1998), 
HD~55496, HD~104340, HD~148897 and HD~206983. Secondary targets, with
their halo classification relying only on the Mennessier et al. (1997)
analysis, are HD~15589, HD~43389 (binary with $P = 1689$~d; 
Jorissen et al. 1998),
HD~123396, HD~139409, HD~187762 and CD$-$27:2233} with intermediate
luminosities, and (ii) their suspected symbiotic activity. 

\subsection{d'-type yellow SyS: young post-PN systems?}

Yellow SyS of type d' (Allen 1982; Schmid \& Nussbaumer 1993) differ
from their s-type counterparts in several respects
(Table~\ref{Tab:yellow}): they host a complex
circumstellar environment (including cool dust, 
bipolar outflows, extended optical nebulae or
emission-line spectra closely resembling those of planetary nebulae), the cool components have early spectral types (F to early
K), they are often
fast rotators (with the possible exception of M 1-2 =V471 Per; Grauer \& Bond
1981) and, finally, they 
belong to the galactic disk unlike s-type yellow SyS which belong to the halo.

All these  arguments  suggest that the hot component in d'-type SyS
just evolved from the AGB to the WD stage. The rather cool dust 
(Schmid \& Nussbaumer 1993) is a relic from the mass lost by the AGB
star. The optical nebulae observed in d'-type SyS are most likely 
genuine planetary nebulae rather than the 
nebulae associated with the ionized wind of the cool component
(Corradi et al. 1999). 
This is especially clear for AS 201
which actually hosts {\it two} nebulae (Schwarz 1991): 
a large fossil planetary nebula detected by direct imaging, and a small
nebula formed in the wind of the current cool component. Finally, the
rapid rotation of the cool component has likely been caused by spin
accretion from the
former AGB wind like in WIRRING systems  (Jeffries \& Stevens
1996; see also Jorissen 2003). The fact that the cool star has
not yet been slowed down by magnetic braking is another indication
that the mass transfer occurred fairly recently (Theuns et al. 1996). 
Finally, one may wonder whether the much
earlier spectral types encountered among yellow d'-type SyS as compared
to s-type SyS bear some relationship to their fast rotation (departure 
from thermal equilibrium?).

\subsection{No extrinsic PRG among red SyS! Why?}

A small number of  galactic red SyS (UV Aur, SS 38, AS 210, HD 59643 = NQ
Gem and V335 Vul) contain
a cool carbon star as cool component, corresponding to a frequency
of 5/176 = 0.03
in the catalogue of Belczy\'nski et al. (2000). 
This small frequency  contrasts with
that  prevailing in the Magellanic Clouds, where 6 out of 11 SyS contain cool
carbon stars (M\"urset, Schild \& Vogel 1996).
The frequency of carbon-rich SyS actually reflects 
the  number ratio of C to M stars in the parent galaxy,  
and this number ratio in
turn reflects the metallicity of the population (Richer 1989).
Therefore, these carbon SyS are likely {\it intrinsic} carbon stars
(M\"urset et al. 1996), i.e., TP-AGB
stars where the carbon observed in the atmosphere results from the
3DUP (see Sect.~\ref{Sect:zoo}).

The question then arises why there are no {\it extrinsic} C or S stars
among SyS (M\"urset \& Schmid 1999), namely cool components polluted by
carbon-rich matter from the  former TP-AGB
companion. Or in other words, why do red SyS
not comply with the {\it binary paradigm} (Sect.~\ref{Sect:zoo})?
There are at least three possible explanations for the fact that 
red SyS contain M rather than S giants:\\
$\bullet$ the hot companion is a main sequence star 
rather than a WD;\\
$\bullet$ the former AGB star did not go through the TP-AGB, i.e., $M_h <
0.5$~\Msun\ (see Sect.~\ref{Sect:s-process});\\
$\bullet$ the former AGB star did go through the TP-AGB, but its  high
metallicity hindered the efficiency of the s-process and of the mass
transfer (see Sect.~\ref{Sect:s-process}).\\

\begin{figure} 
\plotfiddle{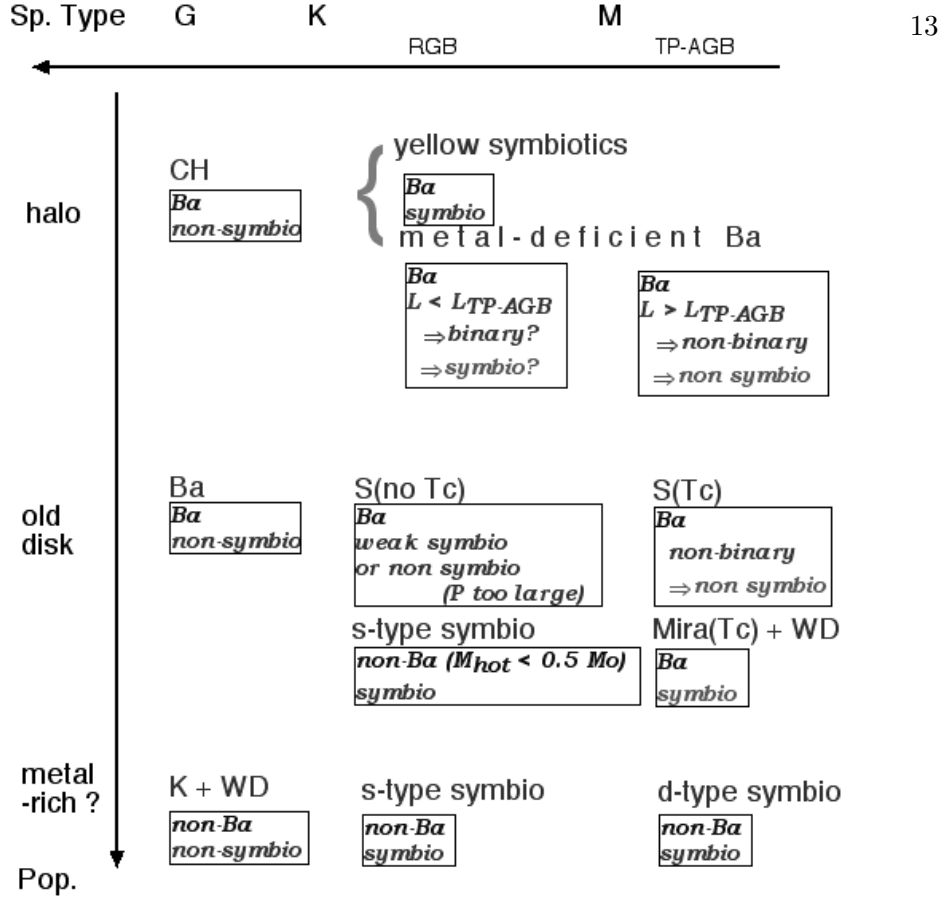}{10cm}{0}{60}{60}{-200}{-80}
\caption[]{\label{Fig:synopsis}
A synopsis of the different families of PRG and SyS stars and their 
relationship 
in terms of presence or absence of symbiotic activity (`symbio') and barium 
syndrome (`Ba').}
\end{figure}

The first question is difficult to address on observational grounds. Let us just
mention here that the eccentricity -- period diagram  may, in some cases,  be used
to  distinguish systems with WD companions from systems with main-sequence
companions. As discussed by Jorissen et al. (1998, their Sect.~ 6 and Fig.~ 4),
binary systems with
$e < 0.1$ and $P > 300$~d most likely host a WD companion, whereas systems
with $ e > 0.23\; ( \log P({\rm  d}) - 1)$ are likely to host main sequence
companions. Since most SyS have nearly circular orbits (Miko\l
ajewska and Hinkle et al., this conference), they are likely to host WD
companions indeed.
 
The second possibility  ($M_h < 0.5$~\Msun) applies to a number of red SyS
with companion masses  fairly accurately determined (see also Miko\l ajewska,
this conference), like  AX Per (0.4~\Msun), EG And ($0.4\pm0.1$~\Msun), SY
Mus ($0.43\pm0.05$~\Msun), RW Hya ($0.48\pm0.06$~\Msun; M\"urset et al.
2000 and references therein).  There are, however, several other red
SyS which do
not fulfill this condition, either marginally (BX Mon: $0.55\pm0.26$~\Msun)  or
more significantly (FG Ser:
$0.60\pm0.15$~\Msun; AR Pav: $0.75\pm0.15$~\Msun; M\" urset et al. 2000;
Schild et al. 2001). For
comparison, the mass of the WD companion in barium systems peaks at 0.67
($\pm0.09$)
\Msun\ (North et al. 2000),  in agreement with the requirement that the AGB
progenitor went through the TP-AGB phase.

Therefore, the third possibility (high $Z$) must be invoked to account for the
lack of barium syndrome in systems like FG Ser or AR Pav  for example, 
which have $M_h > 0.5$~\Msun.
Do red SyS indeed belong to a high-metallicity population?
There are contradictory arguments in that respect. The distribution of carbon
abundances in the cool components of SyS derived by   Schmidt and Miko\l
ajewska (this conference) is representative of red giants having 
slightly subsolar metallicities ([Fe/H] $\sim -0.3$ to $-0.5$). 
On the contrary, Whitelock \&
Munari (1992) showed that the $JHK$ colors of red SyS resemble more the colors
of bulge-like M giants than those 
of normal M giants in the solar neighborhood. 
They argue that this color difference may be related to the higher metallicity of
bulge-like giants, and, hence, of red SyS. A subsequent 
kinematical analysis (Munari 1994) confirmed that   red SyS
belong to the bulge/thick-disk population. 
A direct high-resolution spectroscopic determination of 
the metallicities of red
SyS is needed to definitely settle that question. It must be hoped
that such a study does indeed 
 confirm the expectation of high metallicities for red SyS, otherwise
answers to the lack of barium syndrome different from those discussed
here would have to be found.
 
To conclude, a synopsis  of the different families of PRG and SyS stars, and
their relationship in terms of presence or absence of symbiotic activity
and barium  syndrome, is presented in Fig.~\ref{Fig:synopsis}. 

\acknowledgments
This paper greatly benefited from discussions with H.M. Schmid and
C. Pereira about the nature of d'-SyS.
A.J. is Research Associate of the {\it Fonds National de la Recherche
  Scientifique} (Belgium).

\end{document}